# Some new relativistic charged models for compact objects with anisotropic pressure

Manuel Malaver[1] and Rajan Iyer[2]

[1]Maritime University of the Caribbean, Department of Basic Sciences, Catia la Mar, Venezuela.
 Email: mmf.umc@gmail.com

[2]Environmental Materials Theoretical Physicist, Department of Physical Mathematics Sciences Engineering Project Technologies, Engineeringinc International Operational Teknet Earth Global, Tempe, Arizona, United States of America
 Email: engginc@msn.com

**Abstract:** In this paper, we found new classes of solutions to the Einstein-Maxwell field equations with matter anisotropic distribution incorporating a particular form of electric field intensity within the framework of general relativity. We use a metric potential or ansatz that depends on an adjustable parameter *n* in order to get the new solutions. We generated new models of compact stars with *n*=1 and *n*=2. Graphical analysis allows us to conclude that the new models satisfy all the physical characteristics for astrophysical objects and can be very useful in the study and description of compact structures. We obtained models consistent with the pulsars PSR J1311-3430 and PSR J0952–0607.

## 1. Introduction

Research on compact objects and strange stars within the framework of the general theory of relativity is a central issue of great importance in theoretical astrophysics in the last decades [1,2]. The obtained models in general relativity have been used to describe fluid spheres with strong gravitational fields as is the case in strange stars and neutron stars. The physics of ultrahigh densities is not well understood and many of the strange stars studies have been performed within the framework of the MIT bag model where the matter equation of state has the following linear form $P = \frac{1}{3}(\rho - 4B)$. In this equation $\rho$ is the energy density, $P$ is the isotropic pressure and $B$ is the bag constant. The first detailed models of strange stars based on a strange quark matter equation of state were modeled by Haensel et al. [3] who considered specific features of accretion on strange stars. Also Alcock et al. [4] analyzed strange stars with the normal crust and proposed scenarios for the formation of these compact objects.

Researches as Komathiraj and Maharaj [5], Thirukkanesh and Maharaj [6], Sharma et al. [7], Maharaj et al.[8], Thirukkanesh and Ragel [9,10], Feroze and Siddiqui [11,12], Sunzu et al.[13], Pant et al. [14] and Malaver [15-19], Sunzu and Danford [20], Komathiraj and Maharaj [21] have used numerous mathematical strategies to try to obtain exact solutions which indicates that the Einstein-Maxwell field equations is of great importance to describe compact objects.

The presence of an electric field within a fluid sphere has been a subject of great interest because it has allowed studying the effect of electromagnetic fields on astrophysical stellar objects. According Bhar and Murad [22] the existence of charge affects the values of redshifts, luminosities and mass for stars. Gupta and Maurya [23] have developed some stellar models with a well-defined electric field and Pant et al. [14] studied various solutions for charged matter with finite pressure.

In order to propose physical models of interest that behave well it is important to consider an adequate equation of state. Many researchers have developed exact analytical models of strange stars within the framework of linear equation of state based on MIT bag model together with a particular choice of metric potentials or mass function [24-34]. Mafa Takisa and Maharaj [35] obtained new exact solutions to the Einstein-Maxwell system of equations with a polytropic equation of state. Thirukkanesh and Ragel [36] have obtained particular models of anisotropic fluids with polytropic equation of state which are consistent with the reported experimental observations. Feroze and Siddiqui [11] and Malaver [15] consider a quadratic equation of state for the matter distribution and specify particular forms for the gravitational potential and electric field intensity. Bhar and Murad [22] obtained new relativistic stellar models with a particular type of metric function and a generalized Chaplygin equation of state. Tello-Ortiz et al. [37] also found an anisotropic fluid sphere solution of the Einstein-Maxwell field equations with a modified version of the Chaplygin equation. More recently Malaver and Iyer [38] generated new models of compact stars considering the new version of Chaplygin equation of state proposed for Errehymy and Daoud [39].

In recent decades, the theoretical research [40-50] in realistic stellar models show that the nuclear matter may be locally anisotropic in certain very high density ranges ($\rho > 10^{15}$ gcm$^{-3}$), where the relativistic treatment of nuclear interactions in the stellar matter becomes important. From the pioneering work of Bowers and Liang [40] that generalized the equation of hydrostatic equilibrium for the case local anisotropy, there has been an extensive literature devoted to study the effect of local anisotropy on the bulk properties of spherically symmetric static general relativistic compact objects [51-54]. Therefore, it is always interesting to explore the consequences produced by the appearance of local anisotropy under variety of circumstances.

Presently there are efforts underway to understand the underlying quantum aspects with astrophysical charged Stellar models [55-60].

Shape of the metric potential depends on energy matter quantum wavefunction that can affect local anisotropy with interior criteria to be satisfied by the interior solution as to present a realistic stellar model, especially strange quark stars as well have been key in Quantum Astrophysical projects ongoing, including discontinuum physics [58-64]. There is also study of the quantum particle group theory with authors advancing that will help to classify general field-particle metrics that will match interior spacetime to the exterior spacetime linking also Standard Model and String Theories with Hubble and James Webb Telescope observations of the earlier genesis of galactical stars [59-61,65,66].

The principal motivation of this work is to develop some new analytical relativistic stellar models by obtaining of solutions of Einstein-Maxwell field equations with a linear equation of state with a particular shape of metric potential *Z(x)* dependent on an adjustable parameter *n*. The solutions obtained by satisfying applicable physical boundary conditions provide a mathematically simple family of electrically charged strange stars. The paper is structured as follows: the next section, Sect.2, are presented the interior solutions of Einstein-Maxwell field equations of anisotropic fluid. In Sect. 3, we present the elementary criteria to be satisfied by the interior solution as to present a realistic stellar model. In Sect. 4, physical acceptability conditions are discussed. The interior spacetime will be matched to the exterior spacetime described by the unique Reissner-Nordstrom metric, physically realistic fluid models will be constructed and analysis will be made on the obtained models in Sect. 5. Finally, Sect. 6 discusses and concludes the work.

## 2. The Einstein-Maxwell Field Equations

In this research the interior metric in Schwarzschild coordinates $x^{\mu} = (t, r, \theta, \varphi)$ [68,69] is given in the following form:

$$ds^2 = -e^{2\nu(r)}dt^2 + e^{2\lambda(r)}dr^2 + r^2(d\theta^2 + \sin^2\theta d\varphi^2) \qquad (1)$$

For the metric (1), the Einstein-Maxwell field equations may be expressed as the following system of differential equations [6]:

$$\frac{1}{r^2}(1-e^{-2\lambda}) + \frac{2\lambda'}{r}e^{-2\lambda} = \rho + \frac{1}{2}E^2 \qquad (2)$$

$$-\frac{1}{r^2}(1-e^{-2\lambda}) + \frac{2\nu'}{r}e^{-2\lambda} = p_r - \frac{1}{2}E^2 \qquad (3)$$

$$e^{-2\lambda}\left(\nu'' + \nu'^2 + \frac{\nu'}{r} - \nu'\lambda' - \frac{\lambda'}{r}\right) = p_t + \frac{1}{2}E^2 \qquad (4)$$

$$\sigma = \frac{1}{r^2}e^{-\lambda}(r^2 E)' \qquad (5)$$

where prime (') denotes the derivate with respect to $r$. $\rho$, $p_r$, $p_t$, $\sigma$ denote the energy density, radial pressure, tangential pressure, charge density of the fluid distribution respectively.

The mass of stellar object contained within a sphere of radius r is given by

$$M(r) = \frac{1}{2}\int_0^r \omega^2 \rho(\omega)d\omega \qquad (6)$$

From Eqs. (3) and (4), we can obtain

$$p_t - p_r = e^{-2\lambda}\left(v'' + v'^2 + \frac{v'}{r} - v'\lambda' - \frac{\lambda'}{r}\right) + \frac{1}{r^2}\left(1 - e^{-2\lambda}\right) - \frac{2v'}{r}e^{-2\lambda} - E^2 \qquad (7)$$

With the transformations $x = Cr^2$, $Z(x) = e^{-2\lambda(r)}$ and $A^2 y^2(x) = e^{2v(r)}$ suggested by Durgapal and Bannerji [70], Eqs. (2)-(7) take the following forms:

$$\frac{1-Z}{x} - 2\dot{Z} = \frac{\rho}{C} + \frac{E^2}{2C} \qquad (8)$$

$$4Z\frac{\dot{y}}{y} - \frac{1-Z}{x} = \frac{p_r}{C} - \frac{E^2}{2C} \qquad (9)$$

$$4xZ\frac{\ddot{y}}{y} + (4Z + 2x\dot{Z})\frac{\dot{y}}{y} + \dot{Z} = \frac{p_t}{C} + \frac{E^2}{2C} \qquad (10)$$

$$\frac{\Delta}{C} = 4xZ\frac{\ddot{y}}{y} + \dot{Z}\left(1 + 2x\frac{\dot{y}}{y}\right) + \frac{1-Z}{x} - \frac{E^2}{C} \qquad (11)$$

$$\sigma^2 = \frac{4CZ}{x}\left(x\dot{E} + E\right)^2 \qquad (12)$$

$$M(x) = \frac{1}{4C^{3/2}}\int_0^x \sqrt{\omega}\rho(\omega)d\omega \qquad (13)$$

where dots denotes the derivate with respect to x, $A > 0$ and $C > 0$ are arbitrary constants and $\Delta = p_t - p_r$ is the anisotropic factor which measures the pressure anisotropy within the star. The system of equations (8)-(12) governs the gravitational behaviour for an anisotropic fluid.

In order to obtain physically realistic stellar models, in this paper we assume that the radial pressure and the energy density are related by the following equation:

$$p_r = m\rho \qquad (14)$$

here *m* is a real constant.

## 3. The New Models

To solve the system of equations we take the following potential

$$Z(x) = \frac{1}{(1+ax)^n} \tag{15}$$

$a$ is a real constant and $n$ is an adjustable parameter. The equation (15) for the metric potential is physically realistic because it allows to obtain a monotonic increasing mass function, regular at the centre of the star and also results in deduction of a monotonic decreasing energy density. According Lighuda et al. [71] we also assume for the electric field intensity

$$\frac{E^2}{2C} = kxZ(x) = \frac{kx}{(1+ax)^n} \tag{16}$$

with $k > 0$. This form of electric field gives us a monotonic increasing function, regular at the centre, positive and remains continuous inside of the star. In this paper, we take the specific values of the adjustable parameter as $n=1, 2$.

With $n=1$, by introducing the Eqs. (15)-(16) into the Eq. (8) we obtain.

$$\rho = C\frac{(a-kx)(1+ax)+2a}{(1+ax)^2} \tag{17}$$

and for the Eq. (14) we can written for the radial pressure

$$p_r = mC\frac{(a-kx)(1+ax)+2a}{(1+ax)^2} \tag{18}$$

Substituting $Z$ and Eqs. (16)-(18) in Eq. (9) we obtain.

$$\frac{\dot{y}}{y} = m\frac{(a-kx)(1+ax)+2a}{4(1+ax)} - \frac{kx}{4} + \frac{a}{4} \tag{19}$$

Integrating Eq. (19)

$$y(x) = c_1(1+ax)^{\frac{1}{2}m} e^{\frac{1}{8}x(m+1)(-kx+2a)} \tag{20}$$

where $c_1$ is the constant of integration.

For metric potentials $e^{2\lambda}$ and $e^{2\nu}$ we have, respectively.

$$e^{2\lambda} = 1 + ax \tag{21}$$

$$e^{2\nu} = A^2 c_1^2 (1+ax)^m e^{\frac{x}{4}(m+1)(-kx+2a)} \tag{22}$$

From the Eq. (13), the mass function takes the form

$$M(x) = \frac{\sqrt{x}}{2\sqrt{C}} \left[ \frac{ax}{1+ax} + \frac{k(3-ax)}{3a^2} \right] - \frac{k \arctan(\sqrt{ax})}{2a^2 \sqrt{ac}} \tag{23}$$

Inserting $Z$ and Eq. (16) in Eq. (12) we have for the charge density

$$\sigma^2 = \frac{8C^2 k(ax+2)^2}{(1+ax)^4} \tag{24}$$

From the Eq.(20) and substituting the Eqs.(15) and (16) into the Eq.(11) we obtain for the anisotropy

$$\Delta = \frac{4x}{1+ax} \left[ \begin{array}{c} \frac{(m^2-2m)a^2}{4(ax+1)^2} + \frac{ma}{8(ax+1)}((m+1)(-kx+2a)-kx(m+1)) \\ -\frac{k(m+1)}{4} + \left(\frac{(m+1)(-kx+2a)-kx(m+1)}{8}\right)^2 \end{array} \right]$$
$$-\frac{a}{(1+ax)^2}\left[1+x\left(\frac{ma}{ax+1}+\frac{(m+1)(-kx+2a)-kx(m+1)}{4}\right)\right]+\frac{a}{1+ax}-\frac{2kx}{1+ax} \tag{25}$$

For $n=2$ we can obtain the following analytical model

$$\rho = C \frac{(a^3 - ak)x^2 + (3a^2 - k)x + 6a}{(1+ax)^3} \tag{26}$$

$$p_r = mC \frac{(a^3 - ak)x^2 + (3a^2 - k)x + 6a}{(1+ax)^3} \tag{27}$$

$$e^{2\lambda} = (1+ax)^2 \tag{28}$$

$$y(x) = c_2 (1+ax)^m e^{\frac{1}{8}x(m+1)(a^2 x - kx + 4a)} \tag{29}$$

$$e^{2\nu} = A^2 c_2^2 (1+ax)^{2m} e^{\frac{x}{4}(m+1)(a^2 x - kx + 4a)} \quad (30)$$

$$M(x) = \frac{\sqrt{x}}{2\sqrt{C}} \left[ \frac{ax(ax+2)}{(ax+1)^2} - \frac{2k(ax+1)+k}{2a^2(ax+1)} \right] - \frac{3k \arctan(\sqrt{ax})}{4a^2 \sqrt{ac}} \quad (31)$$

$$\sigma^2 = \frac{2C^2 k (2a^2 x^2 + 3ax + 3)^2}{(1+ax)^6} \quad (32)$$

$$\Delta = \frac{4x}{(1+ax)^2} \left[ \begin{array}{l} \frac{(m^2-m)a^2}{(ax+1)^2} + \frac{ma}{4(ax+1)}\left((m+1)(a^2 x - kx + 4a) + x(m+1)(a^2-k)\right) \\ + \frac{(m+1)(a^2-k)}{4} + \left(\frac{(m+1)(a^2 x - kx + 4a) - x(m+1)(a^2-k)}{8}\right)^2 \end{array} \right]$$

$$- \frac{2a}{(1+ax)^3}\left[1 + 2x\left(\frac{ma}{ax+1} + \frac{(m+1)(a^2 x - kx + 4a) - x(m+1)(a^2-k)}{8}\right)\right] + \frac{a(ax+2)}{(1+ax)^2} - \frac{2kx}{(1+ax)^2}$$

(33)

## 4. Elementary Criteria for Physical Acceptability

A physically acceptable interior solution of the gravitational field equations must comply with the certain physical conditions [36,72]:

(i) The solution should be free from physical and geometric singularities, i.e., $e^\nu > 0$, $e^\lambda > 0$ and $p_r$, $p_t$, $\rho$ are finite in the range $0 \leq r \leq R$.

(ii) The radial and tangential pressures and density are non-negative $p_r$, $p_t$, $\rho \geq 0$

(iii) Radial pressure $p_r$ should be zero at the boundary $r = R$, i.e., $p_r(r=R)=0$, the energy density and tangential pressure may follow $\rho(r=R)=0$ and $p_t(r=R)=0$.

(iv) The condition $0 \leq v_{sr}^2 = \frac{dp_r}{d\rho} \leq 1$ be the condition that the speed of sound $v_{sr}^2$ not exceeds that of light.

(v) Pressure and density should be maximum at the center and monotonically decreasing towards the pressure free interface (i.e., boundary of the fluid sphere). Mathematically $\frac{dp_r}{dr} \leq 0$ and $\frac{d\rho}{dr} \leq 0$ for $0 \leq r \leq R$.

(vi) Pressure anisotropy vanishes at the centre, i.e., $\Delta(0) = 0$ [40].

(vii) The charged interior solution should be matched with the Reissner–Nordström exterior solution, for which the metric is given by:

$$ds^2 = -\left(1 - \frac{2M}{r} + \frac{Q^2}{r^2}\right)dt^2 + \left(1 - \frac{2M}{r} + \frac{Q^2}{r^2}\right)^{-1} dr^2 + r^2\left(d\theta^2 + \sin^2 d\varphi^2\right) \quad (34)$$

through the boundary $r=R$ where $M$ and $Q$ are the total mass and the total charge of the star, respectively.

## 5. Physical Features of the New Models

For each choice of adjustable parameter $n$, we now show physical features of the new models. For the case $n=1$, the metric functions $e^{2\lambda}$ and $e^{2\nu}$ are free from physical and geometric singularities, i.e., $e^{2\nu} > 0$, $e^{2\lambda} > 0$ for $0 \leq r \leq R$ and in $r=0$, $e^{\lambda(0)} = 1$, $e^{2\nu(0)} = A^2 c_1^2$. The radial pressure and energy density are non-negative in the stellar interior and at the boundary $r=R$, $p_r(r=R) = 0$ and $\rho(r=R) \geq 0$. Also, we also have for central density and radial central pressure $\rho = 3aC$, $p_r = 3maC$. As on the surface of the star $p_r(r=R) = 0$ we have

$$R = \frac{\sqrt{2aCk(a^2 - k + \sqrt{a^4 + 10a^2 k + k^2})}}{2aCk} \quad (35)$$

Energy density and radial pressure gradients, $\frac{d\rho}{dr}$ and $\frac{dp_r}{dr}$ respectively, are monotonically decreasing functions with the radial coordinate $r$. For this case we obtain in the range $0 \leq r \leq R$

$$\frac{d\rho}{dr} = -\frac{2kCr(1+aCr^2) - 2aCr(a - kCr^2)}{(1+aCr^2)^2} - \frac{4aCr[(a - kCr^2)(1+aCr^2) + 2a]}{(1+aCr^2)^3} < 0 \quad (36)$$

$$\frac{dp_r}{dr} = -\frac{2kmCr(1+aCr^2) - 2aCr(a - kCr^2)}{(1+aCr^2)^2} - \frac{4amCr[(a - kCr^2)(1+aCr^2) + 2a]}{(1+aCr^2)^3} < 0 \quad (37)$$

From the equations (36) and (37) is deduced that the pressure and density should be maximum at the center and monotonically decreasing towards the surface of the star.

The mass function in $r=R$ can be obtained from the Eq. (23) and we get

$$M(r=R) = \frac{R}{2}\left[\frac{aCR^2}{1+aCR^2} + \frac{k(3-aCR^2)}{3a^2}\right] - \frac{k\arctan(\sqrt{aC}R)}{2a^2\sqrt{aC}} \tag{38}$$

In this model, the radial speed of sound of anisotropic star will be given by

$$0 \leq v_{sr}^2 = \frac{dp_r}{d\rho} = m \leq 1 \tag{39}$$

The interior solution should match continuously with an exterior Reissner-Nordström solution

$$ds^2 = -\left(1-\frac{2M}{r}+\frac{Q^2}{r^2}\right)dt^2 + \left(1-\frac{2M}{r}+\frac{Q^2}{r^2}\right)^{-1}dr^2 + r^2(d\theta^2 + \sin^2 d\varphi^2) \qquad r \geq R$$

This requires the continuity of $e^\nu$, $e^\lambda$ and $q$ across the boundary $r=R$,

$$e^{2\nu} = e^{-2\lambda} = 1 - \frac{2M}{R} + \frac{Q^2}{R^2} \tag{40}$$

where $M$ and $Q$ represent the total mass and charge inside the fluid sphere, respectively. By matching the interior metric function $Z(x)$ with the exterior Reissner-Nordström metric at the boundary $r=R$ we obtain

$$\frac{2M}{R} = \frac{(aC+2kC^2R^2)R^2}{1+aCR^2} \tag{41}$$

With $n=2$, again the metric functions $e^{2\lambda}$ and $e^{2\nu}$ are free from singularities and in the origin $e^{\lambda(0)} = 1$, $e^{2\nu(0)} = A^2 c_2^2$. The radial pressure and the energy density are monotonically decreasing functions and at the boundary $r=R$, $p_r(r=R)=0$ and $\rho(r=R) \geq 0$. In the center $\rho = 6aC$, $p_r = 6amC$ and are positive if $a,m,C > 0$. As $p_r(r=R)=0$ we have

$$R = \frac{\sqrt{2aC(-a^2+k)(3a^2-k+\sqrt{-15a^4+18a^2k+k^2})}}{2aC(-a^2+k)} \tag{42}$$

For the pressure and density gradients within the stellar interior $0 \leq r \leq R$ we obtain

$$\frac{d\rho}{dr} = -\frac{4C^3r^3(-a^3+ak)+2Cr(-3a^2+k)}{(1+aCr^2)^3} - \frac{6aC^4r^5(a^3-ak)+6aC^3r^3(3a^2-k)+36a^2C^2r}{(1+aCr^2)^4} < 0 \tag{43}$$

$$\frac{dp_r}{dr} = -\frac{4mC^3r^3(-a^3+ak)+2Cr(-3a^2+k)}{(1+aCr^2)^3} - \frac{6amC^4r^5(a^3-ak)+6aC^3r^3(3a^2-k)+36a^2C^2r}{(1+aCr^2)^4} \tag{44}$$

Gradients are negative as expected for a realistic star

For the mass function in $r=R$ we have

$$M(r=R) = \frac{R}{2}\left[\frac{aCR^2(aCR^2+2)}{(aCR^2+1)^2} - \frac{2k(aCR^2+1)+k}{2a^2(aCR^2+1)}\right] - \frac{3k\arctan(\sqrt{aC}R)}{4a^2\sqrt{aC}} \tag{45}$$

Also, in this model the condition $0 \leq v_{sr}^2 = \frac{dp_r}{d\rho} = m \leq 1$ is satisfied.

Again, by matching the interior metric function $Z(x)$ with the exterior Reissner-Nordström metric at the boundary $r=R$ we can obtain

$$\frac{2M}{R} = \frac{2aCR^2+(a^2+2k)C^2R^4}{(1+aCR^2)^2} \tag{46}$$

Table 1 shows the values of the chosen physical parameters $k, a, m$ and the masses of the corresponding stellar objects for $n=1$.

**Table 1**. Parameters $k, a, m$ and stellar masses for $n=1$.

| k | a | m | $M(M_\odot)$ |
|---|---|---|---|
| 0.0010 | 0.2 | 1/3 | $4.30M_\odot$ |
| 0.0013 | 0.2 | 1/3 | $4.0\ M_\odot$ |
| 0.0015 | 0.2 | 1/3 | $3.75M_\odot$ |

$M_\odot$ = sun's mass

Figures 1, 2, 3, 4, 5, 6 and 7 represent the plots of $M$, $\frac{E^2}{2C}$, $\sigma^2, \rho, p_r, \Delta$ and $e^{2\nu}$ with the radial parameter for $n=1$. In all the graphs we considered $C=1$.

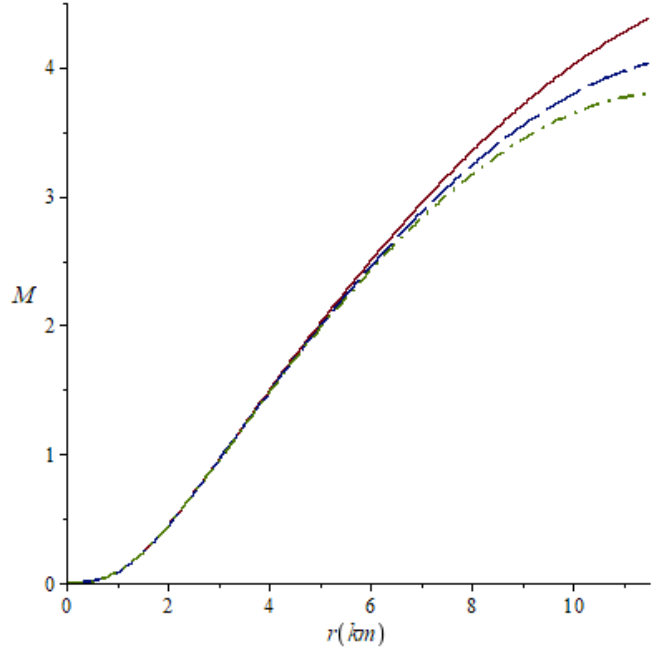

**Figure 1**. Mass function against the radial coordinate for $k=0.0010$ (solid line), $k=0.0013$ (long-dash line) and $k=0.0015$ (dash-dot line). For all the cases $a=0.2$ and $m=1/3$.

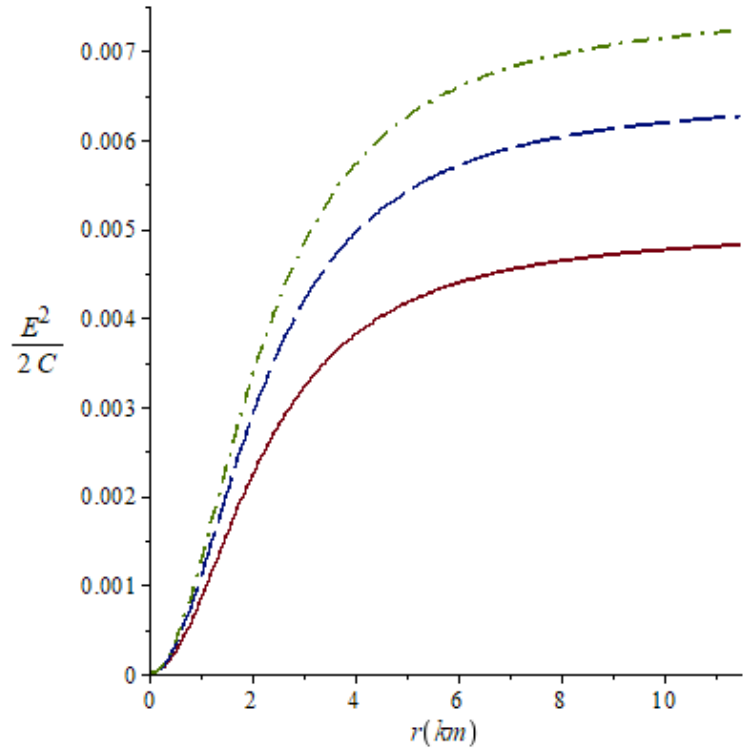

**Figure 2**. Electric field intensity against the radial coordinate for $k=0.0010$ (solid line), $k=0.0013$, (long-dash line) and $k=0.0015$ (dash-dot line). For all the cases $a=0.2$ and $m=1/3$.

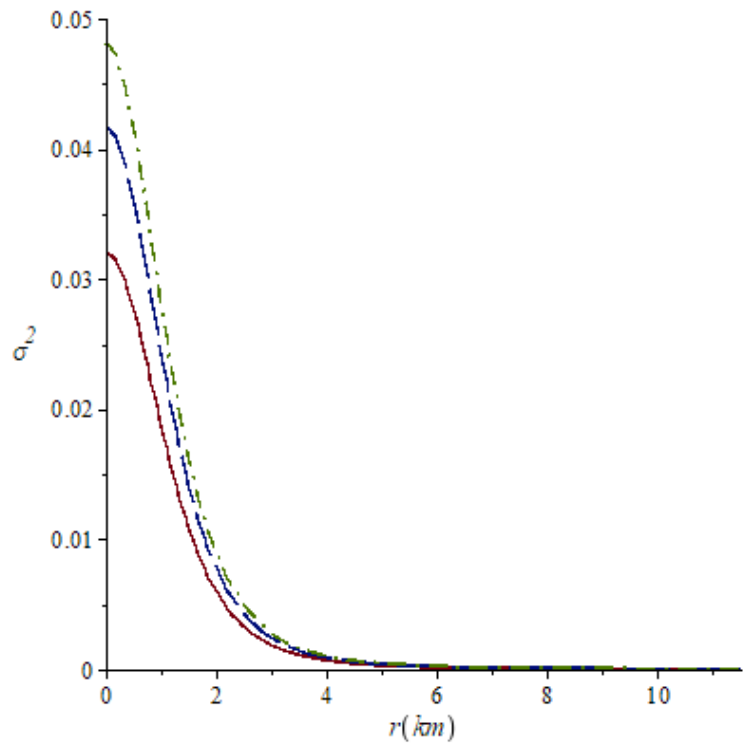

**Figure 3**. Charge density $\sigma^2$ against the radial coordinate for $k=0.0010$ (solid line), $k=0.0013$ (long-dash line) and $k=0.0015$ (dash-dot line). For all the cases $a=0.2$ and $m=1/3$.

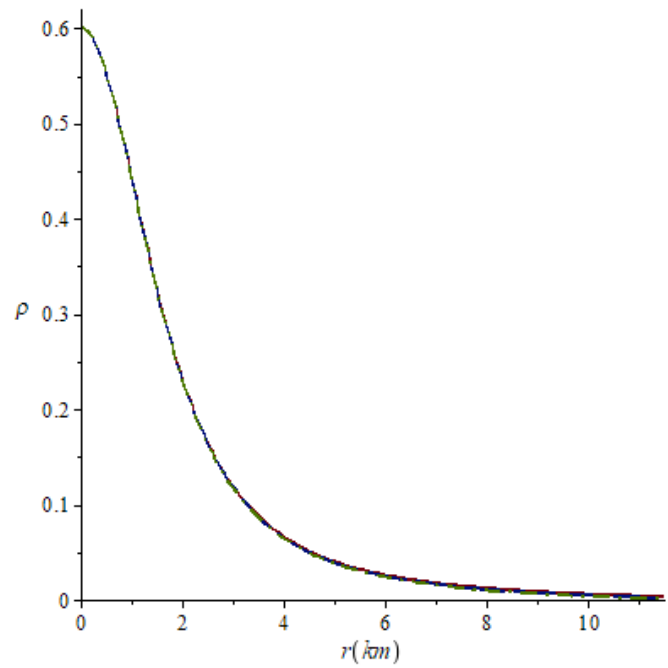

**Figure 4.** Energy density against the radial coordinate for *k*=0.0010 (solid line), *k*=0.0013 (long-dash line) and *k*=0.0015 (dash-dot line). For all the cases *a*=0.2 and *m*=1/3.

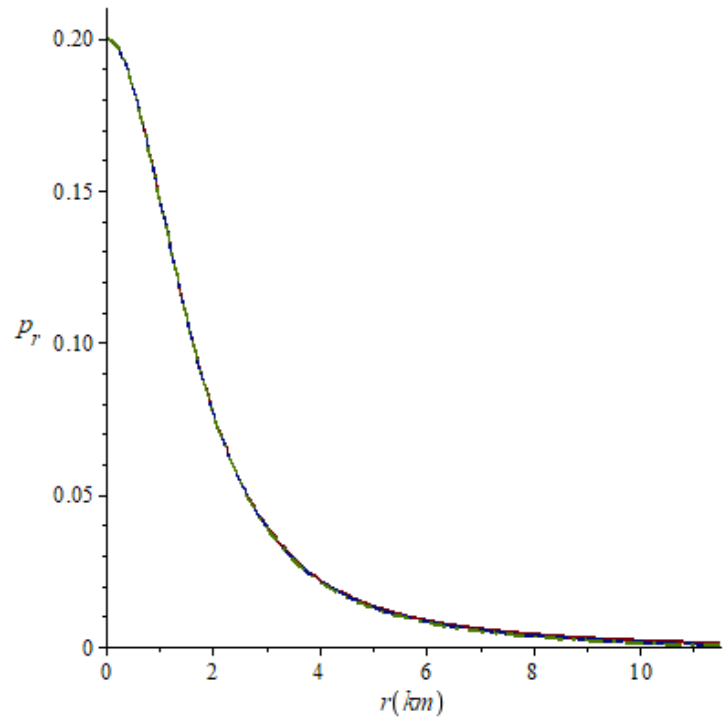

**Figure 5.** Radial pressure against the radial coordinate for *k*=0.0010 (solid line), *k*=0.0013 (long-dash line) and *k*=0.0015 (dash-dot line). For all the cases *a*=0.2 and *m*=1/3.

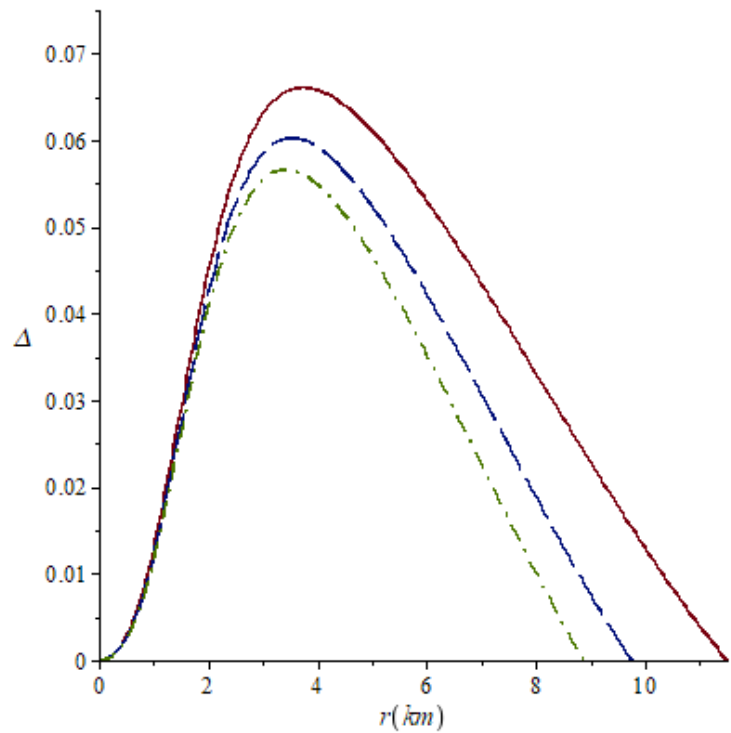

**Figure 6.** Anisotropy **against** the radial coordinate for $k=0.0010$ (solid line), $k=0.0013$ (long-dash line) and $k=0.0015$ (dash-dot line). For all the cases $a=0.2$ and $m=1/3$.

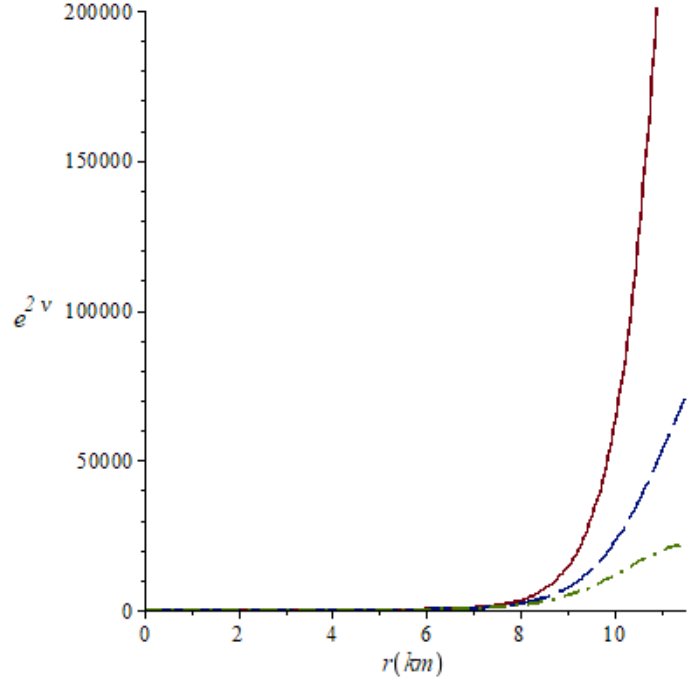

**Figure 7.** Metric potential $e^{2\nu}$ against the radial coordinate for $k=0.0010$ (solid line), $k=0.0013$ (long- dash line) and $k=0.0015$ (dash-dot line). For all the cases $a=0.2$ and $m=1/3$.

Table 2 are shown the values of the chosen physical parameters $k$, $a$, $m$ and the masses of the corresponding stellar objects for $n=2$.

**Table 2**. Parameters $k$, $a$, $m$ and stellar masses for $n=2$.

| k | a | m | $M(M_\odot)$ |
|---|---|---|---|
| 0.0109 | 0.102 | 1/3 | $2.55 M_\odot$ |
| 0.0110 | 0.102 | 1/3 | $2.55 M_\odot$ |
| 0.0111 | 0.102 | 1/3 | $2.55 M_\odot$ |

Figures 8, 9, 10, 11, 12, 13 and 14 represent the plots of $M$, $\dfrac{E^2}{2C}$, $\sigma^2, \rho, p_r, \Delta, e^{2\nu}$ with the radial parameter for $n=2$. Figure 15 shows the variation of the metric function $e^{2\lambda}$ with the radial parameter for $a=0.2$ and $a=0.102$. In all the graphs we considered $C=1$.

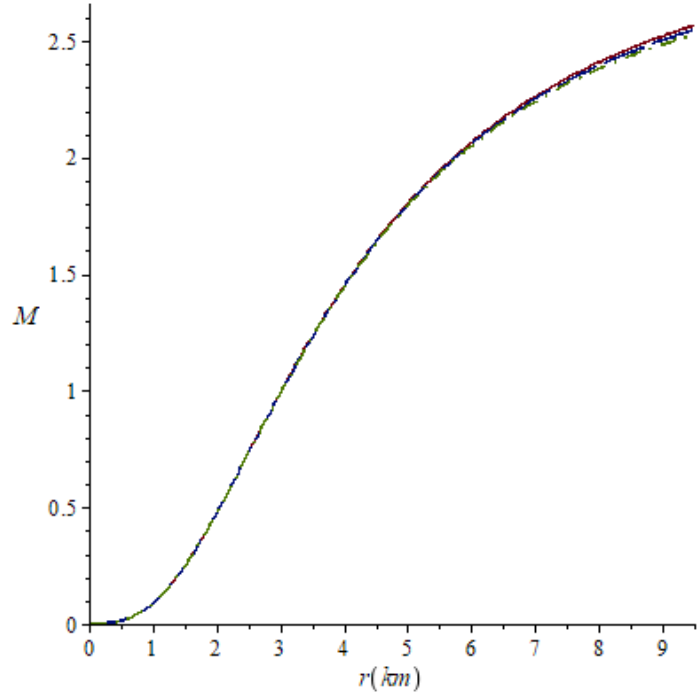

**Figure 8**. Mass function against the radial coordinate for *k*=0.0109 (solid line), *k*=0.011 (long-dash line) and *k*=0.011 (dash-dot line). For all the cases *a*=0.102 and *m*=1/3.

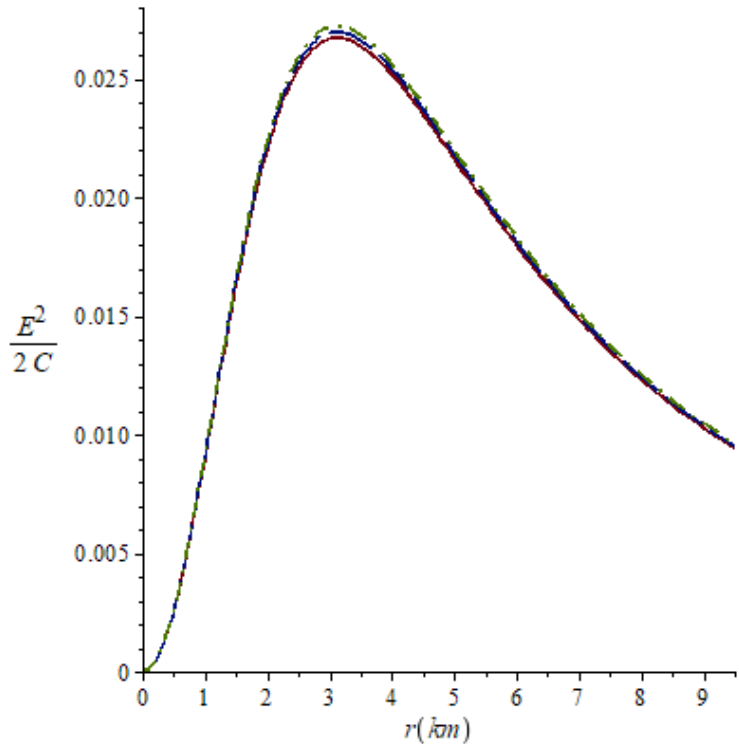

**Figure 9**. Electric field intensity against the radial coordinate for $k=0.0109$ (solid line), $k=0.011$, (long-dash line) and $k=0.0111$ (dash-dot line). For all the cases $a=0.102$ and $m=1/3$.

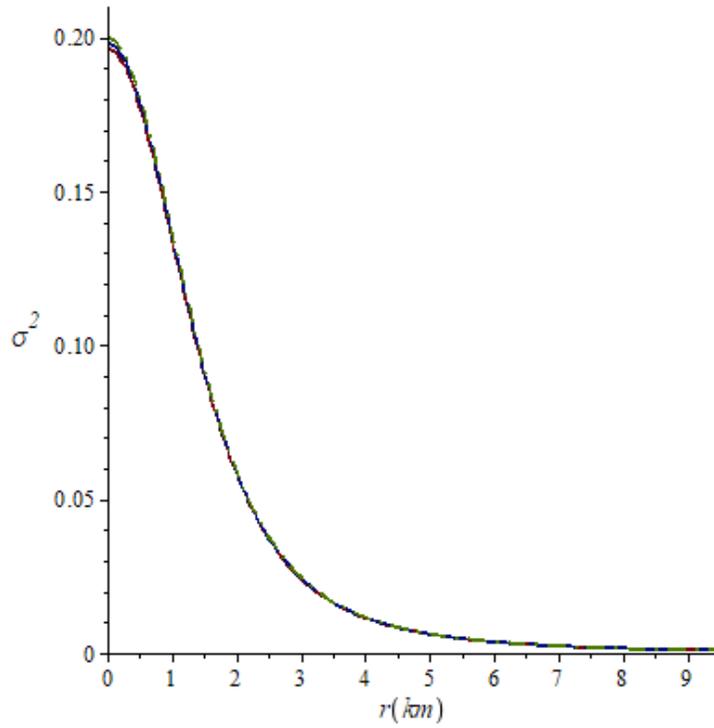

**Figure 10**. Charge density $\sigma^2$ against the radial coordinate for $k=0.0109$ (solid line), $k=0.011$ (long-dash line) and $k=0.0111$ (dash-dot line). For all the cases $a=0.102$ and $m=1/3$.

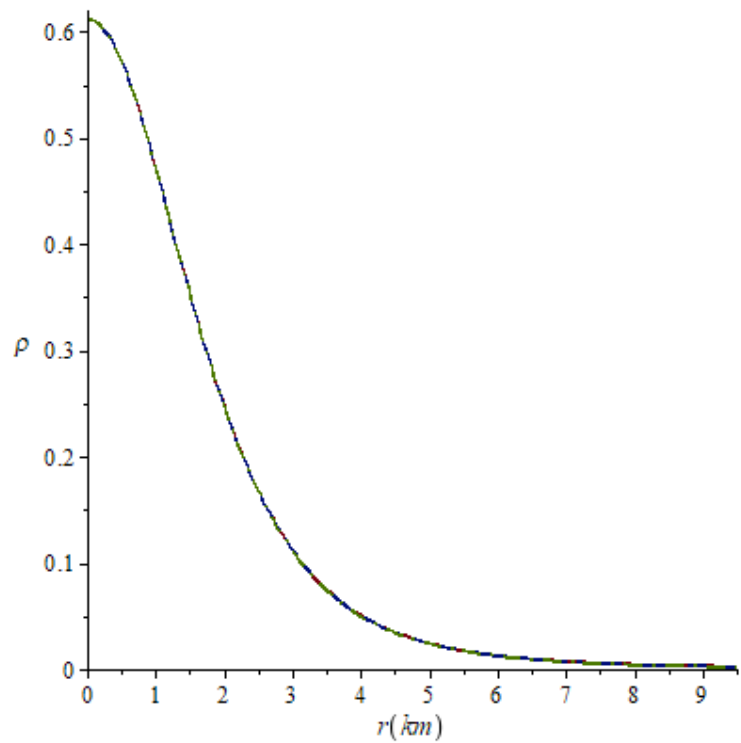

**Figure 11.** Energy density against the radial coordinate for $k=0.0109$ (solid line), $k=0.011$ (long-dash line) and $k=0.0111$ (dash-dot line). For all the cases $a=0.102$ and $m=1/3$.

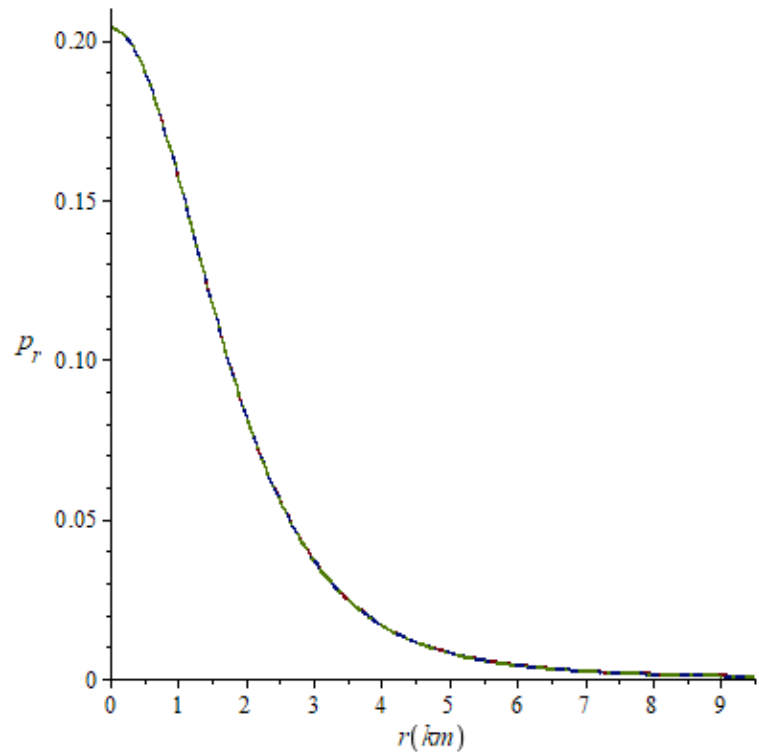

**Figure 12.** Radial pressure against the radial coordinate for $k=0.0109$ (solid line), $k=0.011$ (long-dash line) and $k=0.0111$ (dash-dot line). For all the cases $a=0.102$ and $m=1/3$.

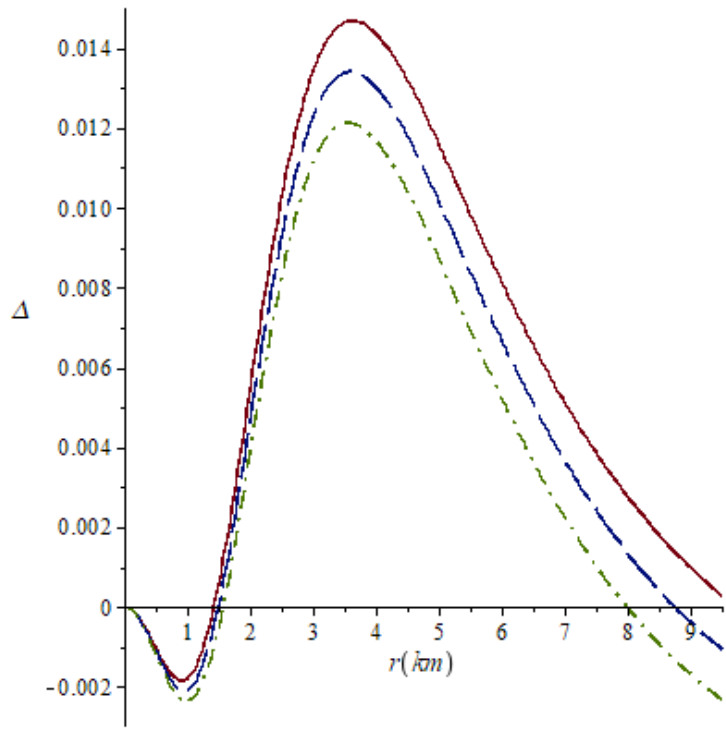

**Figure 13.** Anisotropy against the radial coordinate for k=0.0109 (solid line), k=0.011 (long-dash line) and k=0.0111 (dash-dot line). For all the cases a=0.102 and m=1/3.

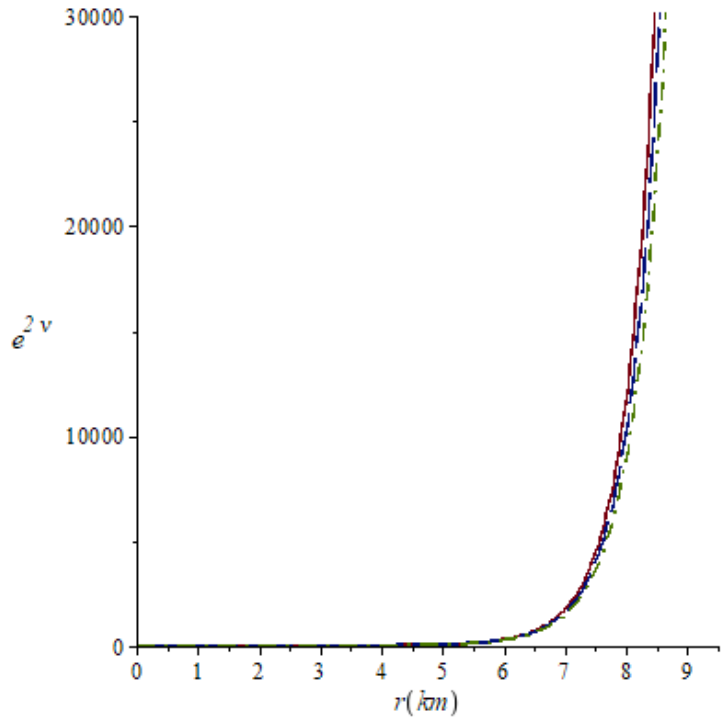

**Figure 14.** Metric potential $e^{2\nu}$ against the radial coordinate for k=0.0109 (solid line), k=0.011 (long- dash line) and k=0.0111 (dash-dot line). For all the cases a=0.102 and m=1/3.

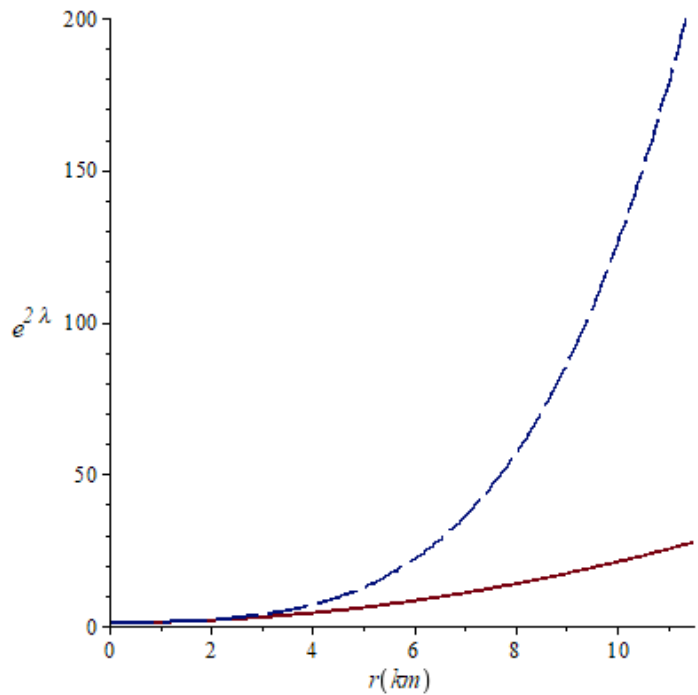

**Figure 15.** Metric potential $e^{2\lambda}$ against the radial coordinate for $a=0.2$ (solid line) and $a=0.102$ (long- dash line).

For the case $n=1$, the behavior of mass function, electric field intensity and charge density with the radial parameter inside the star are presented in Fig.1,2,3, respectively. The Figures show that these physical variables are non-negative, the mass function and electric field are monotonically increasing throughout the fluid distribution while the charge density shows a decrease for all the chosen $k$ values. Higher values of $k$ mean an increase in electric field intensity and charge density, that is $dE/dr > 0$ for $0 \leq r \leq R$, and a decrease in the values associated with the mass function. Figures 4 and 5 is noted that energy density and radial pressure are monotonically decreasing and non-negative functions and not show an appreciable variation with the radial coordinate for the different $k$ values. The variation of anisotropy with the radial parameter is presented in Figure 6. The anisotropy factor initially has an increase, reaches a maximum and then decreases monotonically. High values of $k$ cause a decrease in values of anisotropy with the radial coordinate. In Figure 7, the metric function $e^{2\nu}$ is continuous, well behaved, increases monotonically but shows a decrease when $k$ values increase.

With $n=2$, again the mass function is regular and strictly growing from the center to the surface of the star and not show appreciable variation when $k$ values increase as noted in Figure 8. In Figure 9, the electric field intensity has an initial increase reached a maximum and then decrease monotonically for all values of $k$ and in the Figure 10 is noted that the charge density also is free of singularities, non-negative and decreases with the radial parameter. In Figures 11 and 12 again the energy density and radial pressure have a monotonic decrease with the radial coordinate for the chosen values of k. In Figure 13 the behavior of pressure anisotropy also presents a preliminary growth to reach a maximum and then decrease monotonically. Anisotropy decreases when $k$ values increase. The metric function $e^{2\nu}$ shows continuous growth and well behaved for all $k$ values as noted in Figure 14 and in the Figure 15 the gravitational potential $e^{2\lambda}$ is increased for lower values of $a$ which corresponds to higher values of the parameter $n$, i.e., for $n=2$ we have $a=0.102$ and when $n=1$, $a=0.2$.

We can compare the values calculated for the mass function with observational data. For $n=1$ the values of $k$ and $a$ allow us to obtain an approximate mass of $4M_\odot$ which does not correspond to any known compact star object. With $n=2$, we have obtained masses very similar to the pulsar PSR J1311-3430 with a mass of $2.7M_\odot$ [73] or it could also be related to the pulsar PSR J0952–0607 [74] whose mass is $2.35M_\odot$. Both pulsars are classified as black widow pulsars, which is a type of pulsar hosting a close-orbiting stellar companion that is being consumed by the intense high-energy solar winds of the pulsar and gamma-ray emissions [73-75].

There is also quantum contribution to these mass functions that may shed light on how pulsars close-orbiting stellar companion gets consumed by solar winds and gamma-ray emissions switching quaternion operation of gauge fields of light as well as sound outputs quantum activities [67-75]. The underlying anisotropy mass effects on compact structures like pulsars perhaps will explain their variability with energy density, pressure, mass function, charge density, anisotropy, electric intensity of field, as well as the exterior metric across boundary correlating results observed here; discontiuum physics may be applied to explain consumption by high energy solar winds and gamma-ray emissions [55-67].

## 6. Conclusion

In this work we have developed some simple relativistic charged stellar models obtained by solving Einstein-Maxwell field equations for a static spherically symmetric locally anisotropic fluid distribution. By choice of metric potential and the electric charge distribution together with the linear equation of state we have studied the behavior of fluid distribution. With the positive anisotropy, $p_t > p_r$, the stability of the new solutions is examined by the condition $0 \leq v_{sr}^2 \leq 1$ and it is found that the model developed is potentially stable for the parameters considered. An analytical stellar model with such physical features could play a significant role in the description of internal structure of electrically charged strange stars. The newly obtained models match smoothly with the Schwarzschild exterior metric across the boundary $r=R$ because matter variables and the gravitational potentials of this research are consistent with the physical analysis of these stars.

The new solutions can be related to stellar compact objects such as PSR J1311-3430 [73] and PSR J0952–0607 [74,75]. Physical features associated with the matter, radial pressure, density, anisotropy, charge density and the plots generated suggest that the model with $n=2$ is like the pulsar PSR J0952–0607 and well behaved. We have ansatz formalisms that connect astrophysics with the quantum nature of these anisotropic matter in stellar compact objects, with observable parameters derived from theoretical modeling to experimental measurements. These have all been necessitated by especially current findings of the James Webb Telescope of six earlier formed massive galaxies to peek into quantum nature with our newly developed point-to-point signal/noise matrix measurements of vibrational or sound and photonic or light gauge fields.

Shape of the metric potential depends on energy matter quantum wavefunction that can affect local anisotropy with interior to exterior metric across boundary in compact stellar structures like pulsars. The underlying mass functions have effect on behavior of pressure anisotropy variability alongside energy density, charge density, electric intensity of field, as well as radial metric potential in the interior regions extending to exterior of these stellar objects, consumed by the intense high-energy solar winds of the pulsar and gamma-ray emissions, which are manifesting quantum particle wavefunction astrophysical events, explainable with advancing discontinuum physics with table of realities.